\documentclass[12pt]{article}
\usepackage{latexsym}
\usepackage{graphicx}
\usepackage{cite}
\usepackage{enumitem}
\usepackage{amssymb,mathrsfs}
\usepackage{booktabs}
\usepackage[justification=centering]{caption}
\usepackage{subfigure}
\usepackage{float}

\newtheorem{theorem}{Theorem}[part]
\newtheorem{definition}{Definition}[part]

\newtheorem{remark}{Remark}[part]

\def \ep{\hbox{ }\hfill$\Box$}

\begin{document}
\title{Minimal Integrity Bases of Invariants of Second Order Tensors in a Flat Riemannian Space}

\author{}

\author{Liqun Qi\footnote{%
    Department of Applied Mathematics, The Hong Kong Polytechnic University,
    Hung Hom, Kowloon, Hong Kong ({\tt maqilq@polyu.edu.hk}).
   This author's work was partially supported by the Hong Kong Research Grant Council
   (Grant No. PolyU  15302114, 15300715, 15301716 and 15300717).}
   \and
   Zheng-Hai Huang\footnote{%
    School of Mathematics, Tianjin University, Tianjin 300354, P.R. China ({\tt huangzhenghai@tju.edu.cn}).
   This author was supported by the National Natural Science Foundation of China (Grant No. 11431002).}
   }
\date{\today}

\maketitle

\begin{abstract}
\noindent
In this paper, we study invariants of second order tensors in an $n$-dimensional flat Riemannian space.   We define eigenvalues, eigenvectors and characteristic polynomials for second order tensors in such an $n$-dimensional Riemannian space and show that the coefficients of the characteristic polynomials are real polynomial invariants of that tensor.   Then we give minimal integrity bases for second order symmetric and antisymmetric tensors, respectively, and study their special cases in the Minkowski space and applications in electrodynamics, etc.
\vspace{3mm}

\noindent {\bf Key words:}\hspace{2mm} Flat Riemannian space, second order tensors, eigenvalues, characteristic polynomials, invariants, minimal integrity bases. \vspace{3mm}
%

\end{abstract}

\section{Introduction}
\setcounter{equation}{0} \setcounter{assumption}{0}
\setcounter{theorem}{0} \setcounter{proposition}{0}
\setcounter{corollary}{0} \setcounter{lemma}{0}
\setcounter{definition}{0} \setcounter{remark}{0}
\setcounter{algorithm}{0}

Classical invariant theory was well developed in the nineteenth century.   It was initiated by Gauss \cite{Par-89}.  Cayley \cite{Cri-86, Cri-88} and Hilbert \cite{Hil-93} made fundamental contributions to this development.   In the twentieth century, tensor function representation theory was further developed, and found important applications in theoretical and applied mechanics.  Complete and irreducible polynomial and function representations of vectors, second order tensors and some special fourth order tensors in two- and three-dimensional Euclidean spaces, as well as complete and irreducible tensor function representations involving third order tensors in two-dimensional Euclidean spaces, have been established \cite{Zh-94}.    At the
end of the last century and in the recent years,  minimal integrity bases of isotropic invariants of third and fourth order tensors in the three-dimensional Euclidean spaces were further developed \cite{SB-97, OA-14, OKA-17}.

On the other hand, Einstein's relativity theories were developed with the tool of tensors in the Minkowski and Riemannian spaces \cite{ELMW-52, Rin-06, Bra-04}.   Invariants were also encountered there.   For example, the squares of the norms of the electrical and magnetic field $3$-vectors $\bf e$ and $\bf b$ are no longer invariant in the Minkowski space.   However, the difference of these two squares is an invariant of the electromagnetic field tensor in the Minkowski space \cite{Rin-06, Bra-04}.   The electromagnetic field tensor is a second order antisymmetric tensor in the Minkowski space, while the energy tensor of the electromagnetic field is a symmetric and traceless tensor in the Minkowski space.    May we develop a minimal integrity basis theory for second order tensors in the Minkowski space or a flat Riemannian space?   Such a theory will help us to understand more about invariants in the Minkowski and Riemannian spaces and thus provides a tool for the physics theories in these spaces.

This paper is devoted to this purpose.   We study minimal integrity bases of second order tensors in an $n$-dimensional flat Riemannian space $V$ for $n \ge 2$.

In the next section, we define invariants, polynomial invariants, integrity bases, minimal integrity bases for tensors in $V$.  We show that the number of invariants of each degree in a minimal integrity basis is fixed.

In Section 3, we define powers and traces of second order tensors in $V$.  We show that for a second order symmetric or antisymmetric tensor $A$, for each degree $k \ge 1$, there is only one independent invariant trace$(A^k)$.   Any other invariant of degree $k$ is proportional to  trace$(A^k)$.    Furthermore, we show that trace$(A^k) = 0$ if $k$ is odd and $A$ is antisymmetric.

We define eigenvalues, eigenvectors and characteristic polynomials for second order tensors in $V$ in Section 4.  We show that the coefficients of the characteristic polynomial of a second order tensor $A$ in $V$ are polynomial invariants of $A$.   In this way, we show that if $A$ is symmetric, then $\{ {\rm trace}(A^k): k = 1, \cdots, n  \}$ is a minimal integrity of invariants of $A$, if $A$ is symmetric and traceless, then $\{ {\rm trace}(A^k): k = 2, \cdots, n  \}$ is a minimal integrity of invariants of $A$, and if $A$ is antisymmetric, then $\{ {\rm trace}(A^{2k}): 2 \le 2k \le n, \}$ is a minimal integrity of invariants of $A$.   We also show that the coefficients of their characteristic polynomials also form their minimal integrity bases.

In Section 5, we study the special cases in the Minkowski space and applications in electrodynamics, etc.

\section{Basic Definitions}
\setcounter{equation}{0} \setcounter{assumption}{0}
\setcounter{theorem}{0} \setcounter{proposition}{0}
\setcounter{corollary}{0} \setcounter{lemma}{0}
\setcounter{definition}{0} \setcounter{remark}{0}
\setcounter{algorithm}{0}

Suppose that $V$ is an $n$-dimensional real flat Riemannian space for $n \ge 2$, with a metric tensor $g_{ij}$, where $g_{ij} = g_{ji}$ and det$(g_{ij}) \not = 0$.   The inverse of $g_{ij}$ is $g^{ij}$ such that
$$g^{ij}g_{jk} = \delta^i_{\cdot k},$$
where $\delta^i_{\cdot k}$ is the Kronecker symbol such that
\begin{eqnarray*}
\delta_{\cdot j}^i = \delta^{\cdot j}_i=\left\{\begin{array}{ll}
1 & \mbox{\rm if}\; i=j,\\
0& \mbox{\rm otherwise}.
\end{array}\right.
\end{eqnarray*}

Let $A, \cdots, G$ be tensors in $V$.   Any scalar function $f(A, \cdots, G)$ is called an {\bf invariant} of $A, \cdots, G$.  If $f$ is resulted by tensor operations of $A, \cdots, G$ and some real scalars, then $f$ is called a {\bf polynomial invariant} of $A, \cdots, G$.    If $f$ is resulted by contractions of $A, \cdots, G$, multiplied with some real scalars, where each of $A, \cdots, G$ can be used several times, then $f$ is called a {\bf monomial invariants}.   The degree of a monomial invariant is the times of $A, \cdots, G$ appearing there.  A monomial invariant is called an {\bf irreducible monomial invariant} if it is not a product of monomial invariants of lower degrees.    The sum of monomial invariants with the same degree is a {\bf homogeneous polynomial invariant}.  The sum of irreducible monomial invariants with the same degree is an {\bf irreducible homogeneous polynomial invariant}. A homogeneous polynomial invariant with degree $1$ is called a {\bf linear invariant}.  Then, a polynomial invariant is always the sum of some homogeneous polynomial invariants with different degrees.

A set of polynomial invariants $\{ f_1, \cdots, f_r \}$ of tensors $A, \cdots, G$ is called an integrity basis of $A, \cdots, G$, if any polynomial invariant $f$ is a polynomial of $f_1, \cdots, f_r$.   It is further called a {\bf minimal integrity basis} of $A, \cdots, G$, if none of $f_1, \cdots, f_r$ can be expressed as a polynomial of the other $r-1$ invariants.  According to Hilbert \cite{Zh-94}, $A, \cdots, G$ always have a finite integrity basis.      Also, we may only consider integrity bases consisting of homogeneous polynomial invariants.
Thus, we may talk about degrees of invariants in a minimal integrity basis.

\begin{theorem}\label{basis1}
Suppose that $A, \cdots, G$ are tensors in $V$.  Then for any positive integer $m$, the number of invariants of degree $m$ in any minimal integrity basis of $A,\cdots, G$ is fixed.
\end{theorem}
\noindent {\bf Proof}.   Given a minimal integrity basis $L$ of $A, \cdots, G$, we may always replace each homogeneous polynomial invariant in $L$ by an irreducible homogeneous polynomial invariant with the same degree.      Thus, we may assume that $L$ consists of only irreducible homogeneous polynomial invariants.   Let $M$ be the set of all irreducible homogeneous polynomial invariants of $A, \cdots, G$, with degree $m$.    Then $M$ is a linear space.  By linear algebra, any base of $M$ has the same cardinality, which is the dimension of $M$.    On the other hand, the irreducible homogeneous polynomial invariants of $A, \cdots, G$, in $L$ with degree $m$, forms a base of $M$.  This proves the theorem.
\ep

\section{Traces of Powers of Second Order Tensors in $V$}
\setcounter{equation}{0} \setcounter{assumption}{0}
\setcounter{theorem}{0} \setcounter{proposition}{0}
\setcounter{corollary}{0} \setcounter{lemma}{0}
\setcounter{definition}{0} \setcounter{remark}{0}
\setcounter{algorithm}{0}\setcounter{example}{0}

There are four kinds of forms of second order tensors in $V$: a second order contravariant tensor $A^{ij}$, a second order covariant tensor $A_{ij}$, a second order mixed tensor $A^{\cdot j}_i$, where the covariant index comes first, and a second order mixed tensor $A_{\cdot j}^i$, where the contravariant index comes first.
Their transposes are defined as:
$$(A^{ij})^\top = A^{ji}, \ (A_{ij})^\top = A_{ji}, \ (A^{\cdot j}_i)^\top = A^j_{\cdot i}, \ (A_{\cdot j}^i)^\top =
A_j^{\cdot i}.$$
If
$$A^{ij} = (A^{ij})^\top = A^{ji},$$
then $A^{ij}$ is called a second order symmetric contravariant tensor.   If
$$A^{ij} = -(A^{ij})^\top = -A^{ji},$$
then $A^{ij}$ is called a second order antisymmetric contravariant tensor.   Similarly, if
$$A_{ij} = (A_{ij})^\top = A_{ji},$$
then $A_{ij}$ is called a second order symmetric covariant tensor.   If
$$A_{ij} = -(A_{ij})^\top = -A_{ji},$$
then $A_{ij}$ is called a second order antisymmetric covariant tensor.
There is no second order mixed symmetric or antisymmetric tensor by the definition of the transpose of a second order mixed tensor above.

Note that we may convert these different kinds of second order tensor forms from one to another, such as
$$A^{\cdot j}_i = A_{ik}g^{kj}.$$
Also we see that the contravariant form of a second order tensor is symmetric or antisymmetric if and only if its covariant form is symmetric or antisymmetric respectively.   Thus, we may say that a second order tensor is symmetric or antisymmetric if its contravariant form /covariant form is symmetric or antisymmetric respectively.

The products between vectors and second order tensors follow tensor algebra.   The product of two second order mixed tensors also follow tensor algebra.  In this way, we may define the squares of second order tensors
$$(A^{\cdot j}_i)^2 = A^{\cdot k}_i A^{\cdot j}_k, \ (A_{\cdot j}^i)^2 = A_{\cdot k}^i A_{\cdot j}^k, \
(A^{ij})^2 = A^{ik}g_{kl}A^{lj}, \
(A_{ij})^2 = A_{ik}g^{kl}A_{lj}.$$
We may define higher powers of second order tensors similarly.

The trace of a second order tensor $A^{ij}$ or $A_{ij}$ or $A^{\cdot j}_i$ or $A_{\cdot j}^i$ is defined as
$$A^{\cdot i}_i = A_{\cdot i}^i = A^{ij}g_{ij} = A_{ij}g^{ij}.$$
Hence, we may simply denote it as trace$(A)$.  It is a linear invariant of a second order tensor $A$.   A second order tensor is called a traceless tensor if its trace is equal to zero.  A second order antisymmetric tensor is always traceless as for such a tensor $A$ we have
$${\rm trace}(A) = A^{ij}g_{ij} = -A^{ji}g_{ji} = -{\rm trace}(A).$$

Since
$$A^{ji}g_{ij} = A^{ij}g_{ji} = A^{ij}g_{ij},$$
any linear invariant of a second order tensor $A$ is proportional to its trace.   Then trace$(A^m)$ is a monomial invariant of $A$ with degree $m$.   In general, for $m \ge 2$, there are more than one linearly independent monomial invariants.  For example, in general,
$$A^{ij}g_{jk}A^{lk}g_{li} \not = {\rm trace}(A^2) = A^{ij}g_{jk}A^{kl}g_{li}.$$
However, we have the following theorem.

\begin{theorem}\label{trace}
Suppose that $A$ is a second order symmetric tensor or a second order antisymmetric tensor in $V$.  Then any irreducible homogeneous polynomial invariant of $A$ with degree $m$, where $m$ is positive integer, is proportional to trace$(A^m)$.  If $A$ is antisymmetric and $m$ is odd, then trace$(A^m) = 0$.
\end{theorem}
\noindent {\bf Proof}.    Consider the case that $m=2$.  Let $f$ be an irreducible monomial invariant of $A$, with degree $2$.   Then $f$ is proportional to one of the following three forms:
$${\rm trace}(A^2) = A^{ij}g_{jk}A^{kl}g_{li},$$
$$A^{ij}g_{jk}A^{lk}g_{li}$$
and
$$A^{ji}g_{jk}A^{lk}g_{li}.$$
If $A$ is symmetric, then
$$A^{ij}g_{jk}A^{lk}g_{li} = A^{ji}g_{jk}A^{lk}g_{li} = A^{ij}g_{jk}A^{kl}g_{li} = {\rm trace}(A^2).$$
If $A$ is antisymmetric, then
$$A^{ij}g_{jk}A^{lk}g_{li} = -A^{ij}g_{jk}A^{kl}g_{li} = -{\rm trace}(A^2).$$
and
$$A^{ji}g_{jk}A^{lk}g_{li} = A^{ij}g_{jk}A^{kl}g_{li} = {\rm trace}(A^2).$$
Thus, any irreducible monomial invariant of $A$ with degree $2$ is proportional to ${\rm trace}(A^2)$.  Since an irreducible homogeneous polynomial invariant is a sum of irreducible monomial invariants, it is also proportional to ${\rm trace}(A^2)$.   We may see that the above proof can be extended to $m > 2$.

Now let $m $ be odd and $A$ be antisymmetric.   Then
$${\rm trace}(A^m) = {\rm trace}(A^\top)^m = - {\rm trace}(A^m).$$
Thus,
$${\rm trace}(A^m) = 0.$$
The proof is complete.
\ep

\section{Eigenvalues and Eigenvectors of Second Order Tensors}
\setcounter{equation}{0} \setcounter{assumption}{0}
\setcounter{theorem}{0} \setcounter{proposition}{0}
\setcounter{corollary}{0} \setcounter{lemma}{0}
\setcounter{definition}{0} \setcounter{remark}{0}
\setcounter{algorithm}{0}\setcounter{example}{0}

We now extend $V$ to an $n$-dimensional complex flat Riemannian space $V_C$ such that we may study eigenvalues and eigenvectors of second order tensors in $V$.

\begin{definition}\label{def-eigen}
Consider a second order real mixed tensor $A^i_{\cdot j}$.
If there exist a $\lambda\in \mathbb{C}$ and a nonzero contravariant vector $x^j \in V_C$ such that
\begin{eqnarray}\label{e-eigen-mix1}
A^i_{\cdot j}x^j=\lambda x^i,
\end{eqnarray}
then $\lambda$ is called an eigenvalue of $A^i_{\cdot j}$ and $x^j$ is called an eigenvector associated with the eigenvalue $\lambda$. $(\lambda, x^j)$ is called an eigenpair of $A^i_{\cdot j}$.
\end{definition}

\begin{remark}
 (\ref{e-eigen-mix1}) can be replaced by its contravariant form:
\begin{eqnarray}\label{e-eigen-contravariant}
A^{ij}x_j=\lambda g^{ik}x_k.
\end{eqnarray}
In fact, if (\ref{e-eigen-mix1}) holds, then
$$
A^{ij}x_j=A^{ij}g_{jk}x^k=A^i_{\cdot j}x^j=\lambda x^i=\lambda g^{ij}x_j,
$$
i.e., (\ref{e-eigen-contravariant}) holds; and  if (\ref{e-eigen-contravariant}) holds, then
$$
A^i_{\cdot j}x^j=A^{ik}g_{kj}x^j=A^{ik}x_k=\lambda g^{ik}x_k=\lambda x^i,
$$
i.e., (\ref{e-eigen-mix1}) holds.

Equation (\ref{e-eigen-mix1}) can also be replaced by its mixed form:
$$A^{\cdot j}_i x_j=\lambda x_i,$$
or its covariant form:
$$A_{ij}x^j=\lambda g_{ik}x^k.$$
\end{remark}

The eigenvalue equation (\ref{e-eigen-mix1}) is a tensor equation.   Hence eigenvalue $\lambda$ is an invariant of the second order tensor $A$.    However, the eigenvalue $\lambda$ is not a polynomial invariant of $A$, and may not be real.

From (\ref{e-eigen-mix1}) it follows that
\begin{eqnarray}\label{e-eigen-equa}
(\lambda \delta_{\cdot j}^i-A^i_{\cdot j})x^j=0.
\end{eqnarray}
Since $x^j\neq 0$, it follows from (\ref{e-eigen-equa}) that
$$\phi(\lambda):=\mbox{\rm det}(\lambda \delta_{\cdot j}^i-A^i_{\cdot j})=0.$$
The one dimensional polynomial $\phi(\lambda)$ is called the characteristic polynomial of the second order tensor $A$.
We may write that
$$\phi(\lambda) = \lambda^n + \sum_{k=1}^n (-1)^ka_k(A)\lambda^{n-k}.$$
According to the relationship between the coefficients and roots of a polynomial,  $a_k(A), k = 1, \cdots, n$, are real polynomial invariants of $A$.   In particular,
$$a_1(A) = {\rm trace}(A),$$
and we may define $a_n(A)$ as the determinant of $A$, i.e.,
$$a_n(A) = {\rm det}(A).$$
We see that the Cayley-Hamilton theorem still holds for $A$, i.e.,
$$\phi(A^i_{\cdot j}) \equiv (A^i_{\cdot j})^n + \sum_{k=1}^n (-1)^ka_k(A)(A^i_{\cdot j})^{n-k} = 0^i_{\cdot j},$$
where
$$(A^i_{\cdot j})^0 = \delta^i_{\cdot j}.$$

We now have the following theorem.

\begin{theorem}\label{basis2}
If $A$ is a second order symmetric tensor in $V$, then $\{ {\rm trace}(A^k) : k = 1, \cdots, n \}$ is a minimal integrity basis of $A$, and $\{ a_k(A) : k = 1, \cdots, n \}$ is another minimal integrity basis of $A$.
If $A$ is a second order symmetric and traceless tensor in $V$, then $\{ {\rm trace}(A^k) : k = 2, \cdots, n \}$ is a minimal integrity basis of $A$, and $\{ a_k(A) : k = 2, \cdots, n \}$ is another minimal integrity basis of $A$.
If $A$ is a second order antisymmetric tensor in $V$, then
$$a_k(A) = 0,$$
when $k$ is odd,
$\{ {\rm trace}(A^{2k}) : 2 \le 2k \le n \}$ is a minimal integrity basis of $A$, and $\{ a_k(A) : 2 \le 2k \le n \}$ is another minimal integrity basis of $A$.
\end{theorem}

\noindent {\bf Proof}.
Suppose that $A$ is a second order symmetric tensor in $V$. Let $L_0 = \{ {\rm trace}(A^k) : k = 1, \cdots, n \}$.    Suppose that $f$ is an irreducible homogeneous polynomial invariant of degree $m$.  Then by Theorem \ref{trace}, $f$ is proportional to trace$(A^m)$.  By the Cayley-Hamilton theorem, $A^m$ is a polynomial of $A, A^2, \cdots, A^n$.   Thus, $f$ is a polynomial of the invariants of $L_0$.   This shows that $L_0$ is an integrity basis of $A$.  As ${\rm trace}(A^k)$ for $k = 1, \cdots, n$, are irreducible monomial invariants, we see that $L_0$ is a minimal integrity basis.   It is known that ${\rm trace}(A^k)$ is a polynomial of $a_1(A), \cdots, a_k(A)$ for $k = 1, \cdots, n$ \cite{ZYL-08}, $\{ a_k(A) : k = 1, \cdots, n \}$ is also an integrity basis of $A$.  By Theorem \ref{basis1}, it is also a minimal integrity basis of $A$.

The conclusions for a second order symmetric and traceless tensor follow directly.

We also know that $a_k(A)$ is a polynomial of ${\rm trace}(A), \cdots, {\rm trace}(A^k)$ \cite{ZYL-08}.   If $k$ is odd and $A$ is antisymmetric, by Theorem \ref{trace}, we have $a_k(A) = 0$. The other conclusions can be proved similarly as above.
\ep

\section{The Minkowski Space}
\setcounter{equation}{0} \setcounter{assumption}{0}
\setcounter{theorem}{0} \setcounter{proposition}{0}
\setcounter{corollary}{0} \setcounter{lemma}{0}
\setcounter{definition}{0} \setcounter{remark}{0}
\setcounter{algorithm}{0}\setcounter{example}{0}

Let $V$ be the Minkowski space.  Then $n=4$.   Instead of using $i = 1, 2, 3, 4$, people use $\alpha = 0, 1, 2, 3$ in the Minkwski space, where $\alpha = 0$ corresponds the time component $ct$ with $c$ as the speed of light, and $\alpha = 1, 2, 3$ corresponds the space components.   The metric tensor in the Minkowski space $V$ has the form
$$
G = g^{\alpha\beta}=g_{\alpha\beta}=\left(\begin{array}{cccc}
1 & 0 & 0 & 0\\
0 & -1& 0 & 0\\
0 & 0 &-1 & 0\\
0 & 0 & 0 & -1
\end{array}\right)
=\left(\begin{array}{cc}
1 & {\bf 0}^\top \\
{\bf 0} & -{\bf I}
\end{array}\right),
$$
where $\rm 0$ is the zero $3$-vector, and $\rm I$ is the identity $3$-tensor.

The contravariant form of a second order antisymmetric tensor in $V$ has the form
$$
A^{\alpha\beta}=\left(\begin{array}{cccc}
0 & -e_1 & -e_2 & -e_3\\
e_1 & 0 & -b_3 & b_2\\
e_2 & b_3 & 0 & -b_1\\
e_3 & -b_2 & b_1 & 0
\end{array}\right).
$$

The most well-known example of such a second order antisymmetric tensor is the electromagnetic tensor in electrodynamics \cite{Bra-04, EU-14, Rin-06}.   Then
$$
{\bf e}=\left(\begin{array}{c}
e_1\\
e_2\\
e_3
\end{array}\right)
$$
is the electric field $3$-vector and
$$
{\bf b}=\left(\begin{array}{c}
b_1\\
b_2\\
b_3
\end{array}\right)
$$
is the magnetic field $3$-vector.   The characteristic polynomial of $A^{\alpha\beta}$ is
$$\phi(\lambda) = \lambda^4 + a_2(A)\lambda^2 + a_4(A),$$
where $a_2(A) = -{1 \over 2}{\rm trace}(A^2)$ and $a_4(A) = {\rm det}(A)$.
We have
$$
(A^{\alpha\beta})^2 = A^{\alpha\gamma}g_{\gamma\sigma}A^{\sigma\beta}=\left(\begin{array}{cccc}
e_1^2+e_2^2+e_3^2 & e_2b_3-e_3b_2 & -e_1b_3+e_3b_1 & e_1b_2-e_2b_1\\
e_2b_3-e_3b_2 & -e_1^2+b_3^2+b_2^2 & -e_1e_2-b_1b_2 & -e_1e_3-b_1b_3\\
-e_1b_3+e_3b_1 & -e_1e_2-b_1b_2 & -e_2^2+b_3^2+b_1^2 & -e_2e_3-b_2b_3\\
e_1b_2-e_2b_1 & -e_1e_3-b_1b_3 & -e_2e_3-b_2b_3 & -e_3^2+b_1^2+b_2^2
\end{array}\right),
$$
$$a_2(A) = -{1 \over 2}{\rm trace}(A^2) = -{1 \over 2}A^{\alpha\gamma}g_{\gamma\sigma}A^{\sigma\beta}g_{\beta\alpha} = {\bf b} \cdot {\bf b} - {\bf e} \cdot {\bf e}$$
and
$$a_4(A) = {\rm det}(A) = ({\bf e} \cdot {\bf b})^2.$$
By Theorem \ref{basis2}, $\{  {\bf e} \cdot {\bf e} - {\bf b} \cdot {\bf b}, ({\bf e} \cdot {\bf b})^2 \}$ is a minimal integrity basis of $A^{\alpha\beta}$.  This shows that
${\bf e} \cdot {\bf e}$ and ${\bf b} \cdot {\bf b}$ are not invariants in $V$, but ${\bf e} \cdot {\bf e} - {\bf b} \cdot {\bf b}$ is an invariant in $V$, and it forms a minimal integrity basis of the electromagnetic tensor $A^{\alpha\beta}$, with another invariant $({\bf e} \cdot {\bf b})^2$.   In \cite{Bra-04, EU-14}, ${\bf e} \cdot {\bf b}$ is called a pseudoscalar invariant.   It is invariant under the Lorentz transformations whose determinants are equal to $1$.   In a certain sense, a pseudoscalar invariant is corresponding to a hemitropic invariant in \cite{Zh-94}.

Another example of a second order antisymmetric tensor in the Minkowski space is the acceleration tensor in covariant theory of gravitation \cite{Fed-16}.   Then $\bf e$ and $\bf b$ here are the acceleration field strength $3$-vector and the solenoidal acceleration $3$-vector.

The contravariant form of a second order symmetric tensor in $V$ has the form
$$
A^{\alpha\beta}=\left(\begin{array}{cc}
d & {\rm p}^\top \\
{\rm p} & -{\rm T}
\end{array}\right),
$$
where $d$ is the $(0, 0)$ component of $A$, $\rm p$ is a $3$-vector, and $\bf T$ is a symmetric $3$-tensor.

The most well-known example of a second order symmetric and traceless tensor $A$ is the electromagnetic stress-energy tensor in electrodynamics \cite{Bra-04, Rin-06}.  In this case, $d$ is the energy density and $\bf p$ is the Poynting vector, and ${\rm T}$ is the Maxwell stress tensor.   In this case, we have $a_1(A)=$ trace$(A)=0$.   The characteristic polynomial of $A^{\alpha\beta}$ is
$$\phi(\lambda) = \lambda^4 + a_2(A)\lambda^2 - a_3(A)\lambda^3 + a_4(A),$$
where
$$a_2(A) = -{1 \over 2}{\rm trace}(A^2) = -{1 \over 2}\left[d^2 -2{\rm p}^\top{\rm p} + {\rm trace}({\rm T}^2)\right],$$
$$a_3(A) ={1 \over 3}{\rm trace}(A^3)= {1 \over 3}\left[d^3-3d{\rm p}^\top{\rm p} - 3{\rm p}^\top{\rm T}{\rm p} + {\rm trace}({\rm T}^3) \right]$$
and
\begin{eqnarray*}
a_4(A) &=& {\rm det}(A)\\
&=& {1 \over 8}\left[{\rm trace}(A^2)\right]^2 - {1 \over 4}{\rm trace}(A^4)\\
&=& -d^4 + 12 {\rm p}^\top{\rm p}-4d^2{\rm p}^\top{\rm p} -2 ({\rm p}^\top{\rm p})^2 + ({\rm trace}({\rm T}^2))^2\\
& & +2d^2{\rm trace}({\rm T}^2)-4{\rm p}^\top{\rm p}\cdot {\rm trace}({\rm T}^2)+8d{\rm p}^\top{\rm T}{\rm p}+4{\rm p}^\top{\rm T}^2{\rm p}-{\rm trace}({\rm T}^4).
\end{eqnarray*}
We also have
$$d = {\rm trace}(A) + {\rm trace}({\rm T}) = {\rm trace}({\rm T}),$$
which is always nonnegative.

An example of a second order symmetric tensor is the stress-energy tensor in gravitation theory \cite{MTW-73}.   In this case, we have
$$a_1(A) = {\rm trace}(A) = d - {\rm trace}({\rm T}),$$
\begin{eqnarray*}
a_2(A) &=& {1 \over 2}\left[({\rm trace}(A))^2 - {\rm trace}(A^2)\right]\\
&=& {1 \over 2}\left[2{\rm p}^\top{\rm p}-2d\cdot {\rm trace}({\rm T})+({\rm trace}({\rm T}))^2-{\rm trace}({\rm T}^2)\right],
\end{eqnarray*}
\begin{eqnarray*}
a_3(A) &=& {1 \over 6}\left[({\rm trace}(A)^3 -3{\rm trace}(A){\rm trace}(A^2)+2{\rm trace}(A^3) \right]\\
&=&{1 \over 6}\left[-3d({\rm trace}({\rm T}))^2-({\rm trace}({\rm T}))^3-3d{\rm trace}({\rm T}^2)+6{\rm p}^\top{\rm p}\cdot {\rm trace}({\rm T})\right.\\
& &\left.-3{\rm trace}({\rm T}){\rm trace}({\rm T}^2)-6{\rm p}^\top{\rm T}{\rm p}+2{\rm trace}({\rm T}^3)\right]
\end{eqnarray*}
and
\begin{eqnarray*}
a_4(A) &=& {\rm det}(A) \\
&=& {1 \over 24}({\rm trace}(A))^4+{3 \over 8}{\rm trace}(A){\rm trace}(A^3) - {1 \over 4}\left[{\rm trace}(A)\right]^2{\rm trace}(A^2)\\
& & + {1 \over 8}\left[{\rm trace}(A^2)\right]^2 - {1 \over 4}{\rm trace}(A^4)\\
&=&{1 \over 24}\left[4d({\rm trace}({\rm T}))^3+({\rm trace}({\rm T}))^4-12d\cdot {\rm trace}({\rm T}){\rm trace}({\rm T}^2)\right.\\
& &\quad +12{\rm p}^\top{\rm p}({\rm trace}({\rm T}))^2-6({\rm trace}({\rm T}))^2{\rm trace}({\rm T}^2)+8d\cdot {\rm trace}({\rm T}^3)\\
& &\quad -24{\rm p}^\top{\rm T}{\rm p}\cdot {\rm trace}({\rm T})+8{\rm trace}({\rm T}){\rm trace}({\rm T}^3)+3({\rm trace}({\rm T}^2))^2-12d^2{\rm p}^\top{\rm p}\\
& &\quad \left.+24{\rm p}^\top{\rm p}-12{\rm p}^\top{\rm p}\cdot {\rm trace}({\rm T}^2)-12({\rm p}^\top{\rm p})^2+24{\rm p}^\top{\rm T}^2{\rm p}-6{\rm trace}({\rm T}^4)\right].
\end{eqnarray*}
We do not go to more details.


\end{document}